\documentstyle[amssymb,11pt]{article}

\newtheorem{Theo}{Theorem}     
\newtheorem{Prop}{Proposition} 
\newtheorem{Lem}{Lemma}

\def\traitfinal{\nobreak\bigskip\nobreak%
\hbox to\textwidth{\hfil\hbox to 8 cm{\hrulefill}\hfil}}

\def\Abs#1{\left\vert #1 \right\vert}   

\def\CV_#1^#2{\mathrel{
\mathop{\kern 0pt\hbox to 10mm{\rightarrowfill}}
\limits_{#1}^{#2}}}

\def\E{\mathop{\hbox{E}}\nolimits}              
\def\Var{\mathop{\hbox{Var}}\nolimits}          

\def\Eqalign#1{\null\,\vcenter{\openup\jot\m@th\ialign{
\strut\hfil$\displaystyle{##}$&$\displaystyle{{}##}$\hfil
&&\quad\strut\hfil$\displaystyle{##}$&$\displaystyle{{##}}$
\hfil\crcr#1\crcr}}\,}

\def\system#1{\left\{\null\,\vcenter{\openup1\jot\m@th%
\ialign{\strut\hfil$##$&$##$\hfil&&\enspace$##$\enspace&%
\hfil$##$&$##$\hfil\crcr#1\crcr}}\right.}

\newcommand{\Cz}{h_n=\petito{k_n}}
\newcommand{\Cu}{k_n=\petito{n/\ln n}}
\newcommand{\Cd}{n=\petito{k_n^{1+\alpha}}}

\newcommand{\somi}{\sum_{i=0}^{h_n}}
\newcommand{\somr}{\sum_{r=1}^{k_n}}

\newcommand{\xnret}{X_{n,r}^{\star}}
\newcommand{\znret}{Z_{n,r}^{\star}}

\newcommand{\ynt}{Y_{n,t}}
\newcommand{\ynr}{Y_{n,r}}
\newcommand{\yns}{Y_{n,s}}
\newcommand{\unr}{u_{n,r}}

\newcommand{\inr}{I_{n,r}}
\newcommand{\Fnr}{F_{n,r}}
\newcommand{\Fnrb}{\bar{F}_{n,r}}
\newcommand{\xinr}{\xi_{n,r}}
\newcommand{\phinr}{\Phi_{n,r}}
\newcommand{\anr}{a_{n,r}}
\newcommand{\enr}{\varepsilon_{n,r}}
\newcommand{\Mnr}{M_{n,r}}
\newcommand{\Dnr}{D_{n,r}}

\newcommand{\mnr}{m_{n,r}}
\newcommand{\lnr}{\lambda_{n,r}}

\newcommand{\aiknchap}{\hat a_{i,k_n}}
\newcommand{\fnchap}{\hat f_n}
\newcommand{\fnchapx}{\hat f_n(x)}

\newcommand{\proof}{{\bf Proof~: }}
\newcommand{\CQFD}
{%
\mbox{}%
\nolinebreak%
\hfill%
\rule{2mm}{2mm}%
\medbreak%
\par%
}
\newcommand{\proba}[1]
{P\left(\left\{{#1}\right\}\right)}
\newcommand{\normsup}[1]
{{\parallel{#1}\parallel}_{\infty}} 
 
\newcommand{\normdeux}[1]
{{\parallel{#1}\parallel}_2} 
\newcommand{\grando}[1]
{O\left({#1}\right)} 
\newcommand{\petito}[1]
{o\left({#1}\right)} 
\newcommand{\indic}[1] 
{{\bf 1}_{#1}}

\begin{document}
\title{Extreme values and Haar series estimates\\ of point process boundaries}
\author{St\'ephane Girard \& Pierre Jacob\\\\
Universit\'e Montpellier 2}
\date{}
\maketitle

\begin{abstract}
We present a new method for estimating the edge of a two-dimensional
bounded set, given a finite random set of points drawn from the interior.
The estimator is based both on Haar series and extreme values of the point process.
We give conditions for various kind of convergence and we obtain remarkably different
possible limit distributions. We propose a method of reducing the negative bias,
illustrated by a simulation.

\end{abstract}

\section{Introduction}

Many proposals are given in the literature for estimating a bounded set
$S$ of ${\mathbb R}^2$, given a finite random set $N$ of points drawn from the interior.
Such a diversity follows from crossing properties of the observed random set $N$
(sample, point process, random field on a grid), properties of the unknown
bounded set $S$ (convex sets, star-shaped domains, pieces of support under a given
curve, images), with the relevant treatments (convex hulls, extreme values,
functional estimates, change-point estimates, segmentation).

Here $N$ is a Poisson point process, with observed points belonging to a subset 
$S=\{(x,y):\;~0\leq~x\leq~1\;~;~\;~0\leq~y~\leq~f(x)\}$, where $f$ is an unknown
function, so that the problem reduces to estimating $f$.
This problem arises for instance in econometrics where the function $f$ is
called the production frontier.
This is the case in H\"ardle {\it et al.}~(1995b)
where data consist of pairs $(X_i,Y_i)$,
$X_i$ representing the input (labor, energy or capital) used to produce
an output $Y_i$ in a given firm $i$. In such a framework, the value $f(x)$
 can be interpreted as the maximum level of output which is attainable 
for the level of input $x$.

An early paper was written by Geffroy~(1964) for samples where the estimator
was a kind of histogram based on extreme values.
More precisely, taking the equidistant partition of $[0,1]$ into $k_n$
intervals, this estimator $\hat{g}_n$ is defined by choosing on each interval
the uppest observed point.
 Some developments were achieved by Chevalier~(1976), Gensbittel~(1979),
 and Jacob~(1984). Estimating a smooth star-shaped contour is a
closely related matter which was adressed by Jacob and Abbar~(1989),
Abbar~(1990) and de Haan and Resnick~(1994) with extreme values methods. 
In Deprins {\it et al.}~(1984) and more recently, in Korostelev~(1995), $f$ is 
from economical considerations supposed to be
increasing and convex which suggests an adapted
estimator, called the DEA (Data Envelopment Analysis) estimator.
 Asymptotic minimax optimality of piecewise polynomials estimates is studied
by Korostelev and Tsybakov~(1993), Mammen and Tsybakov~(1995), Gayraud~(1997)
and H\"ardle {\it et al.}~(1995a).
Let us note that a precise knowledge of the regularity of the function 
subject to estimation is a necessary condition to obtain this optimality.
If it is only known that $f$ is a $C^1$ function, this estimator does not differ
from the estimator $\hat{g}_n$ of Geffroy.

The first attempt in the direction of orthogonal
series was done by Abbar and Suquet~(1993), and more thoroughly by
Jacob and Suquet~(1995).
The method, due to Cencov~(1962) and Tiago de Oliveira (1963), 
consists of estimating the
first coefficients of a truncated $L^2-$ expansion $f_n$ of $f$.
The case of the Haar basis, the general case of $C^1$ bases,
and the example of the trigonometric 
basis were investigated, in the case of Poisson process with a known intensity $c$,
permitting the use of averaging estimates of the Fourier coefficients. In the present
paper, the intensity is supposed to be unknown, and we present extreme-valued
estimates of these coefficients in the case of the Haar basis. 
Then, our estimates
$\fnchap$ will appear as a linear combination of a set of $k_n$ extreme values,
involving the Dirichlet kernel associated to the $(h_n+1)$ first Haar functions.
It is convenient to consider $k_n$ as an integer multiple of $(h_n+1)$.
In the case $(h_n+1)=k_n$, $\fnchap$ turns out to be simply the estimator $\hat{g}_n$
of Geffroy. In the general case $k_n=d_n(h_n+1)$, with $d_n\to\infty$.
This is one of the advantages of $\fnchap$ to be simple to compute. Moreover,
it does not require either the knowledge of the intensity of the point process
or information on the regularity or geometry of the function.

Of course, it is understood in the sequel that $h_n\to\infty$ so that
$k_n\to\infty$, whatever the choice of $d_n$. In sections \ref{secbiais} and 
\ref{seccons}, we produce mild conditions for the convergence of $\fnchap$,
in terms of local error, mean integrated square error, 
and for the uniform norm.  In subsection \ref{localcv},
the choice of $d_n=1$ or $d_n\to\infty$ appears mainly as a conflict between
the systematic part of the bias, $f-f_n$, and the variance $\Var{\fnchap}$,
while the statistical part of the bias, $\E{\fnchap}-f_n$, remains unaffected.
The choice $d_n=1$ privileges the systematic bias, but the choice $d_n\to\infty$
improves the variance. In the case $d_n=1$, the limit distribution of $\fnchap$
is extremal, while in the case $d_n\to\infty$, $\fnchap$ is asymptotically normal,
with a better speed of convergence.

In fact, the most awkward part of the bias, in either method, is a negative
component of the statistical part $\E{\fnchap}-f_n$, which is of order $k_n/(nc)$.
Recall that $c$ is assumed to be unknown. In section \ref{secred},
we present a very simple
and efficient method of eliminating this component. The resulting improvement
is quite evident on simulations, see Girard and Jacob (2003). What is better, we obtain in this way the
asymptotical distribution for $\fnchap-f$ instead of $\fnchap-\E{\fnchap}$.

\section{Preliminaries}
\label{prelimin}

Let $N$ be a stationary Poisson point process on ${\mathbb R}^2$ with an unknown
intensity rate $c>0$. Actually, suppose that we merely observe the truncated
point process $N(.\cap S)$ where
\begin{equation}
\label{defS}
S=\{(x,y)\in {\mathbb R}^2\mid 0\leq x\leq 1~;\;0\leq y \leq f(x)\},
\end{equation}
and where $f$ is a measurable function such that
\begin{equation}
\label{mM}
0<m=\inf_{[0,1]}f \leq M=\sup_{[0,1]}f<+\infty.
\end{equation}
For each positive integer $i$, let us denote by $J_i$ the interval
$J_i=\left[\frac{p_i}{ 2^{q_i-1}},\frac{p_i+1}{ 2^{q_i-1}}\right)$ where
$p_i$ and $q_i$
are the integers uniquely determined by
$i= 2^{q_i-1}+p_i$ and $0\leq p_i < 2^{q_i-1}$. For $i=2^{q_i}-1$, let us
close $J_i$ on the right.
In what follows we suppose that $h_n+1=2^{h'_n}$, $h'_n\in{\mathbb N}$,
so that $\{J_\ell$, $h_n+1\leq\ell\leq 2h_n+1\}$ appears as a partition of the unit
interval into $(h_n+1)$ equidistant intervals. For each $x\in[0,1]$, denote
$J_\ell(x)$ the interval that contains $x$.
Let  $(k_n)$ be a further increasing sequence of integers such that 
$k_n=d_n(h_n+1)$, where $d_n\in{\mathbb N}$, $d_n>0$.
Take the equidistant partition of the unit interval into $k_n$ intervals $\inr$,
$r=1,\dots,k_n$ centered at $x_r=(2r-1)/(2k_n)$, and the associated partition
of $S$ into $k_n$ cells $D_{n,r}$ with a basis $\inr$:
\begin{equation}
D_{n,r}= \{\;(x,y) \in S \mid x\in I_{n,r}\;\}.
\end{equation} 
Then, writing $\lnr$ for the Lebesgue measure $\lambda (D_{n,r})$, the
piecewise constant function
\begin{equation}
\label{serieHaar}
f_n(x)=(h_n+1) \sum_{x_r\in J_\ell(x)}  \lnr, \qquad x\in [0,1]
\end{equation}
appears as an approximation of $f$.\\
Let $N^{*n}$ denote the sum of independent Poisson processes with the same intensity
rate $c$. $N^{*n}$ can be considered as a Poisson process with mean measure $nc\lambda$.
For $r=1,\dots,k_n$, let $\xnret=0$ if $N^{*n}(D_{n,r})=0$. If $N^{*n}(D_{n,r})>0$,
let $\xnret$ denote the supremum of the second coordinate of the points of the
truncated process $N^{*n}(.\cap D_{n,r})$.
We estimate each $\lnr$ by $\xnret/k_n$, thus
obtaining an estimator $\fnchap$ of $f_n$ which is
\begin{equation}
\label{estimateur}
\fnchapx= \frac{1}{d_n} \sum_{x_r\in J_\ell(x)}  \xnret, \qquad x\in [0,1].
\end{equation}
For the sake of brevity, we write, for $r=1,\dots,k_n$,
$
\ynr=({\xnret}/{k_n})-\lnr,
$
obtaining the reduced formula
\begin{equation}
\label{reduced}
\fnchapx-f_n(x)  = (h_n+1) \sum_{x_r\in J_\ell(x)} \ynr, \qquad x\in [0,1].
\end{equation}
In a first reading of our paper, this matter-of-fact presentation
of $\fnchap$ is sufficient. However, the deep motivation of this definition
is that $\fnchap$ belongs to the important family of orthogonal series
estimators. Moreover, in most proofs, $\fnchap$ is easy to handle in this
setting.\\
As a square integrable function on $[0,1]$, $f$ admits a $L^2$-convergent series
expansion with respect to every orthonormal basis of $L^2([0,1])$. 
In particular, the Haar basis is defined by
\begin{equation}
\label{defbase}
e_0=\indic{[0,1]}, \qquad
e_i=2^{\frac{q_i-1}{2}}\left(\indic{J_{2i}}-\indic{J_{2i+1}}\right),
\quad i\geq 1.
\end{equation}
The Dirichlet kernel of order $h_n$ associated to the Haar basis is given by:
\begin{equation}\label{noyauHaar}
K_n(x,y)=\somi e_i(x)e_i(y)=(h_n+1) \indic{J_\ell(x)}(y).
\end{equation}
Now, let $\sum_{i=0}^{\infty}a_i e_i$ be the $L^2$-expansion of $f$,
 where $a_i=\int_0^1 e_i(t)f(t)dt$. The truncated expansion
$\somi a_i e_i(x)$, $x\in [0,1]$
can be rewritten in terms of the Dirichlet kernel as
$\int_0^1 K_n(x,y)f(y) dy$.
In view of (\ref{serieHaar}) and (\ref{noyauHaar}), we obtain:
\begin{equation}
\label{trunc}
\somi a_i e_i(x)=\int_0^1 K_n(x,y)f(y) dy=(h_n+1)\int_{J_\ell(x)} f(y)dy=f_n(x),\; x\in[0,1].
\end{equation}
Note also that $\fnchapx$ can be written on the Haar basis as 
\begin{equation}
\fnchapx= \sum_{i=0}^{h_n} \aiknchap e_i(x), \qquad x\in [0,1],
\end{equation}
where $\aiknchap$ is an estimate of the coefficient $a_i$ defined by
\begin{equation}
\label{defaiknchap}
\aiknchap=\somr e_i(x_r) \frac{\xnret}{k_n}, \qquad 0\leq i\leq h_n.
\end{equation}
A straightforward calculation leads to the formula
\begin{equation}
\label{difference}
\fnchapx = \somr K_n(x_r,x) \frac{\xnret}{k_n}, \qquad x\in [0,1],
\end{equation}
which provides a general expression for orthogonal series estimates of $f$.
Before proceeding, we collect some further notations to be used in the sequel:
$$
\min_{x\in\inr} f(x) = \mnr,\; \sup_{x\in\inr} f(x) = \Mnr.
$$

\section{Systematic bias convergence}
\label{secbiais}

According to section 2, $\fnchap$ is an estimate of the approximation $f_n$ of the
unknown function $f$. Thus, a preliminary examination of the convergence of the
systematic bias $f_n-f$ is required.
We need the modulus of continuity of $f$:
$$
\omega(f,\delta)=\sup\{\Abs{f(x+\eta)-f(x)};\;0\leq x\leq x+\eta\leq 1;\;
 0\leq\eta<\delta \}.
$$
Then,
\begin{equation}
\Abs{f_n(x)-f(x)} \leq (h_n+1) \int_{J_{\ell}(x)} \Abs{f(x)-f(y)}dy \leq
\omega(f,1/(h_n+1)).
\end{equation}
In view of the last inequality, the following proposition is clear:
\begin{Prop}
\label{propcu}
If $f$ is $\alpha$-Lipschitzian $(0<\alpha\leq 1)$,
then $\normsup{f_n-f}=\grando{h_n^{-\alpha}}$.
\end{Prop}
The convergence of $\normdeux{f_n-f}$ is easily deduced from
Proposition~\ref{propcu} since $\normdeux{f_n-f}^2\leq \normsup{f_n-f}^2$.
\begin{Prop}
\label{propcvL2}
If $f$ is $\alpha$-Lipschitzian $(0<\alpha\leq 1)$, 
then $\normdeux{f_n-f}^2=\grando{h_n^{-2\alpha}}$.
\end{Prop}
Notice that, from (\ref{trunc}), $f_n-f$ is merely the remainder of the partial sum of order $h_n$ of the Haar series of $f$, so that
$h_n=\petito{n}$ is a sufficient condition for both the $L^2$-convergence
and the uniform convergence when $f$ is continuous.

\section{Consistency results}
\label{seccons}

For $r=1,\dots,k_n$, the truncated Poisson point processes $N^{*n}(.\cap D_{n,r})$
are independent, and the extreme values $\xnret$ inherit this property. Thus,
the distribution of the estimator $\fnchap$ is uniquely determined by the
distribution of each $\xnret$.
We refer to the Appendix for a series of lemmas describing these distributions.

\subsection{Local convergence }
\label{localcv}

\begin{Theo}
\label{thlocal}
Suppose $f$ satisfies a Lipschitz condition of order $\alpha$, $(0<\alpha\leq 1)$.
Then, under the condition $\Cu$, we have for all $x\in[0,1]$:
\begin{equation}
\Abs{\E{\fnchapx}-f_n(x) + \frac{k_n}{nc} }=\grando{\frac{1}{k_n^\alpha}}.
\end{equation}
If, moreover, $\Cd$~then 
\begin{equation}
\Var{\fnchapx} \sim \frac{k_n h_n}{n^2c^2}.
\end{equation}
\end{Theo}
The condition $\Cd$ is necessary to control the variance of $\xnret$ 
(see Lemma \ref{lemespe}~(\ref{troisi})).
It imposes that, for all $r=1,\dots,k_n$,
the mean number of points
above $\mnr$ in $\Dnr$ converges to~0. It is then possible
to ignore the contribution of the events $\{\xnret> \mnr\}$ to the variance.

\noindent
\proof
In view of Lemma \ref{lemespe}~(\ref{deuxi}),
\begin{equation}
\max_{1\leq r\leq k_n}\Abs{\E\left(\xnret\right)-\mnr+\frac{k_n}{nc}}=
\grando{\frac{1}{k_n^\alpha}}.
\end{equation}
Thus, since $f$ is $\alpha$-Lipschitzian,
\begin{equation}
\Abs{\E{\fnchapx}-f_n(x) + \frac{k_n}{nc} } = \grando{\frac{1}{k_n^\alpha}}.
\end{equation}
Moreover, in view of Lemma \ref{lemespe}~(\ref{troisi}),
\begin{equation}
\Var{\fnchapx} = \Var{\left((h_n+1)\sum_{x_r\in J_\ell(x)} \frac{\xnret}{k_n}\right)}
= \frac{1}{d_n^2}\sum_{x_r\in J_\ell(x)} \Var{\xnret} \sim \frac{k_n h_n}{n^2c^2}.
\end{equation}
\CQFD
\noindent Of course, for $d_n=1$, the variance is the worst since
$k_n h_n /(n^2 c^2)\sim k_n^2/(n^2 c^2)$. (See the discussion on the choice
of $d_n$ in the introduction.).

\subsection{Mean integrated square convergence }

\begin{Theo}
\label{thmisc}
Suppose $f$ satisfies a Lipschitz condition of order $\alpha$, $(0<\alpha\leq 1)$.
Then, for the mean integrated square convergence of $\fnchap$ to $f$ it is
sufficient that $\Cu$. 
\end{Theo}
Let us note that the condition $\Cu$~is necessary for the application of Lemma \ref{lemespe}~(\ref{uni}).

\noindent \proof
Utilizing Fubini's theorem, the mean integrated square error can be written as
\begin{equation}
J(\fnchap)=\int_0^1 \E(\fnchapx - f(x))^2\,dx = \E\left(\normdeux {\fnchap - f}^2\right).
\end{equation}
Taking account of the $L^2$-orthogonality of $\fnchap-f_n$ and $f_n-f$, we obtain
\begin{equation}
J(\fnchap)=\E\left(\normdeux {\fnchap - f_n}^2\right) + \normdeux {f_n - f}^2,
\end{equation}
where the second term goes to zero by Proposition~\ref{propcvL2}.
In view of (\ref{reduced}), the first term can be expanded as
\begin{equation}
\E\left(\normdeux {\fnchap - f_n}^2\right)= (h_n+1) \sum_{\ell=h_n+1}^{2h_n+1}
\E\left[\left(\sum_{x_r\in J_\ell(x)} \ynr\right)^2\right].
\end{equation}
From the inequality
\begin{equation} 
\E(\ynr\yns) \leq \frac{1}{2} ( \E(\ynr^2) + \E(\yns^2)) \leq \max_{1\leq t\leq k_n} \E(\ynt^2),
\end{equation}
we obtain
\begin{equation} 
\E\left(\normdeux {\fnchap - f_n}^2\right)\leq k_n^2 \max_{1\leq t\leq k_n}\E(\ynt^2),
\end{equation}
and the result follows from Lemma \ref{lemespe}~(\ref{uni}):
\begin{equation}
\label{eqJ}
\E\left(\normdeux{\fnchap-f_n}^2\right)=\grando{\frac{k_n^2}{n^2}}+\grando{\frac{1}{k_n^{2\alpha}}}.
\end{equation}
\CQFD

\subsection{Almost complete uniform convergence  }

We shall give sufficient conditions for the convergence of the series
\begin{equation}
\forall \varepsilon >0, \qquad \sum_{n=1}^{+\infty}\proba{\normsup{\fnchap-f}>\varepsilon }<+\infty.
\end{equation}
\begin{Theo}
\label{thacuc}
If $f$ is continuous, then under the condition $\Cu$,
$\fnchap$ converges to $f$ almost completely uniformly.
\end{Theo}
\proof
The condition $\Cu$ implies $h_n=\petito{n}$ and consequently the
continuity of $f$ yields the uniform convergence of $f_n$ to $f$. 
 Thus, it remains to consider $\fnchap -f_n$, which is given by (\ref{reduced}).
Then, we have
\begin{equation}
\normsup{\fnchap-f_n}=(h_n+1) \max_{h_n+1\leq \ell \leq 2h_n+1}
\Abs{\sum_{x_r\in J_{\ell}}\ynr},
\end{equation}
which implies
\begin{eqnarray}
\proba{\normsup{\fnchap-f_n}>\varepsilon}
& \leq & \sum_{\ell =h_n+1}^{2h_n+1}\proba{\Abs{\sum_{x_r\in J_{\ell}}\ynr}>\frac{\varepsilon}{h_n+1}}, \nonumber \\
& \leq & \sum_{\ell =h_n+1}^{2h_n+1}\proba{\bigcup_{x_r\in J_{\ell}}\left\{|\ynr|>\frac{\varepsilon}{d_n (h_n+1)}\right\}},
\end{eqnarray}
where $d_n=k_n/(h_n+1)$ represents the cardinality of $\{x_r\in J_{\ell}\}$. Then,
\begin{equation}
\proba{\normsup{\fnchap-f_n}>\varepsilon} \leq \sum_{\ell =h_n+1}^{2h_n+1}
\sum_{x_r\in J_{\ell}} \proba{\Abs{\xnret-k_n\lnr}> \varepsilon}.
\end{equation}
Since $f$ is continuous, for $n$ large enough $\Mnr-\mnr<\varepsilon/2$ uniformly in $r$, and therefore 
$
\left\{\Abs{\xnret-k_n\lnr}> \varepsilon\right\} \subset \left\{\xnret<\mnr-\varepsilon/2\right\}. 
$
This yields
\begin{equation}
\proba{\normsup{\fnchap-f_n}>\varepsilon} \leq \sum_{\ell =h_n+1}^{2h_n+1}
\sum_{x_r\in J_{\ell}} \Fnr(\mnr-\varepsilon/2).
\end{equation}
In view of Lemma \ref{lemfdr}, $\Fnr(\mnr-\varepsilon/2)$ can be computed,
leading to
\begin{equation}
\proba{\normsup{\fnchap-f_n}>\varepsilon} \leq \sum_{\ell =h_n+1}^{2h_n+1}
\sum_{x_r\in J_{\ell}} \exp{\left(-\frac{nc\varepsilon}{2 k_n}\right)}
\leq  k_n \exp{\left(-\frac{nc\varepsilon}{2 k_n}\right)},
\end{equation}
which is the general term of a convergent series provided $\Cu$.
\CQFD

\subsection{Asymptotic distribution}

It is worth noticing that in convergence Theorems \ref{thlocal}, \ref{thmisc},
and \ref{thacuc}, conditions $\Cu$~and $\Cd$~do not imply that 
$k_n$ should go to infinity faster than $h_n$.
 On the contrary, the asymptotic distribution of $\fnchap$
depends strongly on the relative rates of $h_n$ and $k_n$. We present two
entirely different cases. In the first one ($d_n=1)$, the partition
$\{J_{\ell},\; h_n+1\leq \ell\leq 2h_n+1\}$, is arranged to be exactly the same as the partition
$\{\inr,\; 1\leq r \leq k_n\}$, so that the estimator $\fnchap$ merely reduces to $\hat{g}_n$, the Geffroy's estimate, a kind of histogram based
on the uppermost values of the point process on each $\inr$. Such a case has already been
extensively studied since the generating paper of Geffroy~(1964) and gives extremal
asymptotic distributions.
The second case ($d_n\to\infty$) is more in the spirit of the orthogonal series method, with a condition
$h_n=\petito{k_n}$ suggesting an estimation of the coordinates $a_i=\int_0^1 e_i(t)f(t)dt$
by the random Riemann sums $\aiknchap$, see (\ref{defaiknchap}). Due to the underlying properties
of the Poisson process $N^{*n}$, these $\aiknchap$ are sums of independent terms, so that a
Gaussian asymptotic distribution is expected. However, in both theorems, a closer control of $k_n$
is required, which is essentially of the same kind as $n=\petito{k_n^{1+\alpha}}$.
\begin{Theo}
\label{thweib}
Let $x\in[0,1]$ and suppose $f$ is $\alpha$-Lipschitzian $(0<\alpha\leq 1)$. 
Suppose $d_n=1$ and $\Cd$.
\begin{enumerate}
\item [(i)] If $k_n=\petito{n}$, then $T_n(x)=\frac{nc}{k_n}(\fnchapx-k_n \lnr)$ converges in distribution to the Weibull Extremal Value Distribution.
\item [(ii)] Suppose $k_n\ln k_n=\petito{n}$ and denote $Z_n=\displaystyle \max_{1\leq r\leq k_n} \sup_{x \in \inr} |\fnchap(x)-\Mnr|$, then
$\frac{nc}{k_n} Z_n-\ln{k_n}$ converges in distribution to the Gumbel 
Extremal Value Distribution.
\end{enumerate}
\end{Theo}
Let us note that in (ii), condition $k_n\ln k_n=\petito{n}$ is just a weaker
version of the condition $\Cu$.
\noindent
\proof 
\begin{enumerate}
\item [(i)]
Denote by $G_{n,x}$ the distribution function of $T_n(x)$, and consider $r$ such
that $x\in \inr$. Then for all $u\in {\mathbb R}$,
\begin{equation}
G_{n,x}(u)  = \proba{\fnchapx \leq k_n\lnr + u k_n/nc}=\Fnr(k_n\lnr + u k_n/nc),
\end{equation}
since $\fnchapx$ reduces to $\xnret$ on $\inr$ when $d_n=1$. 
Suppose for instance that $u\leq 0$. Then $k_n=\petito{n}$
implies $n\lnr\to\infty$ and thus for $n$ large enough,
$$
u\in \left[-nc\lnr,nc\left(\frac{\Mnr}{k_n}-\lnr\right)\right].
$$
Lemma \ref{lemfdr} yields
$
G_{n,x}(u) = \exp{(u-nc\xinr(u))}
$
with 
\begin{equation}
\xinr(u)=\enr(k_n\lnr + u k_n/nc)-\lnr+\Mnr/k_n,
\end{equation}
and $0\leq nc \xinr(u) \leq nc (\Mnr\!\!-\!\mnr)/k_n$.
Then $\xinr\to 0$ as $n\to\+\infty$ under condition $\Cd$.
As a conclusion, $G_{n,x}(u)\to \exp(u)$ if $u\leq 0$.
The case $u>0$ is similar, the result being $G_{n,x}(u)\to 1$.
\item [(ii)]
When $d_n=1$, $\fnchapx$ reduces to $\xnret$ on $\inr$, and thus
$Z_n=\displaystyle \max_{1\leq r\leq k_n} (\Mnr - \xnret)$.
Let $G_n$ stand for the distribution function of $\frac{nc}{k_n} Z_n-\ln{k_n}$.
Consider $u\in{\mathbb R}$.
 For $n$ large enough
 $$
u\in \left[-\ln k_n,\frac{nc}{k_n}\left(\Mnr-\frac{k_n\ln k_n}{nc}\right)\right],
$$
 $\forall r\in\{1,\dots,k_n\}$ and by a straightforward computation
\begin{equation}
G_n(u) = \prod_{r=1}^{k_n} \Fnrb\left(\Mnr -\frac{k_n}{nc}(u+\ln{k_n})\right)
 =  \prod_{r=1}^{k_n} \left(1- e^{-\eta_{n,r}}\frac{e^{-u}}{k_n}\right),
\end{equation}
with $nc(\lnr-\Mnr/k_n)\leq \eta_{n,r} \leq 0$.
Now, $\displaystyle \prod_{r=1}^{k_n} \left(1- \frac{e^{-u}}{k_n}\right)\to e^{-e^{-u}}$,
$\eta_{n,r}$ is a null array in view of $\Cd$, and Lemma \ref{lemmnull} (see Appendix) gives the result.
\CQFD
\end{enumerate}
\noindent Now, let $x \in [0,1]$ be fixed. Denote ${\sigma}_n = k_n/(nc d_n^{1/2})$. 

\begin{Theo}
\label{thnorasymp1}
If $f$ is $\alpha$-Lipschitzian $(0<\alpha\leq 1)$, under the three conditions 
$\Cz$, $\Cu$ and $n=\petito{k_n^{1/2+\alpha} h_n^{1/2}}$,
$V_n(x)={\sigma}_n^{-1}(\fnchapx - \E(\fnchapx))$
converges in distribution to a standard Gaussian variable.
\end{Theo}
\proof
Let us denote by $\psi_{n,x}$ the characteristic function of $V_n(x)$.
It expands as
\begin{eqnarray}
\label{decomp2}
\psi_{n,x}(t) & = & \exp{it\left(\frac{nc }{k_n d_n^{1/2}}\sum_{x_r\in J_{\ell}(x)} 
(\mnr-\E(\xnret)) - d_n^{1/2} \right)} \nonumber \\
&\times& \E\left[\exp{ it\left(\frac{nc }{k_n d_n^{1/2}}\sum_{x_r\in J_{\ell}(x)}
(\xnret - \mnr) + d_n^{1/2} \right)}\right].
\end{eqnarray}
Introducing $\anr=\mnr-k_n/nc$, (\ref{decomp2}) can be rewritten as
\begin{eqnarray}
\psi_{n,x}(t) & = & \exp{it\left(\frac{nc }{k_n d_n^{1/2}}\sum_{x_r\in J_{\ell}(x)} 
(\anr-\E(\xnret))\right) } \nonumber \\
&\times& \E\left[\exp{ it\left(\frac{nc }{k_n d_n^{1/2}}\sum_{x_r\in J_{\ell}(x)}
(\xnret - \mnr) +  d_n^{1/2} \right)}\right].
\end{eqnarray} 
In view of Lemma \ref{lemespe}~(\ref{deuxi}), the first term in the product converges to $1$ under conditions $\Cz$, $\Cu$ and $n=\petito{k_n^{1/2+\alpha} h_n^{1/2}}$.
After simplification, and in view of the independence of the $\xnret$, 
\begin{equation}
\psi_{n,x}(t)  \sim  e^{it d_n^{1/2}}
\prod_{x_r\in J_{\ell}(x)}\E\left[\exp{ it\left(\frac{nc }{k_n d_n^{1/2}}
(\xnret - \mnr) \right)}\right].
\end{equation}
Introducing the characteristic function of $\xnret-\mnr$, it follows that
\begin{equation}
\psi_{n,x}(t)  \sim  e^{it d_n^{1/2}} \prod_{x_r\in J_{\ell}(x)} \phinr(t_n)
\end{equation}
where $t_n=tnc d_n^{-1/2}/ k_n$.
Before applying Lemma \ref{lemphi}, it remains to verify that $t_n=\petito{n/k_n}$ and
$t_n=\petito{k_n^{1+2\alpha}/n}$.
Clearly,
\begin{equation}
\label{maj1}
t_n\frac{k_n}{nc}=   t d_n^{-1/2}  =\petito{1}
\end{equation}
since $\Cz$, and
\begin{equation}
\label{maj2}
t_n\frac{n}{k_n^{1+2\alpha}} =  tc d_n^{-1/2} \left(\frac{n}{k_n^{1+\alpha}}\right)^2=\petito{1},
\end{equation}
since $\Cz$ and $\Cd$. The characteristic function is of the order:
\begin{eqnarray}
\label{part3}
\psi_{n,x}(t) & \sim &  \exp{\left(it d_n^{1/2}\right)}
 \left( 1+it d_n^{-1/2}\right )^{- d_n}  \\
& \times & 
\prod_{x_r\in J_{\ell}(x)}
\left( 1 + it d_n^{-1/2} \Fnrb(\mnr) +
\grando{\Abs{t_n}\frac{n}{k_n^{1+2\alpha}}} + \petito{n^{-s}} \right).
\label{part4}
\end{eqnarray}
By a third-order Taylor expansion, one sees that (\ref{part3}) converges to
$e^{-t^2/2}$ as $d_n\to+\infty$.
Finally, consider the logarithm of (\ref{part4}):
\begin{equation}
W_n=\sum_{x_r\in J_{\ell}(x)} \ln{(1+\unr)}, \mbox{ with } \unr= it d_n^{-1/2} \Fnrb(\mnr) +
\grando{\Abs{t_n}\frac{n}{k_n^{1+2\alpha}}} + \petito{n^{-s}}. 
\end{equation}
Observe that $\max_{1\leq r\leq k_n} \Abs{\unr}$ converges to 0 with (\ref{maj1}) and (\ref{maj2}).
Thus, for $n$ large enough $\Abs{\unr}~<~1/2$ uniformly in $r$ and the
classical inequality $\Abs{\ln(1+\unr)}<2\Abs{\unr}$ yields
\begin{equation}
\Abs{W_n}\leq 2\Abs{t} d_n^{1/2} \Fnrb(\mnr) + \grando{\Abs{t_n} d_n\frac{n}{k_n^{1+2\alpha}}} + \petito{d_n n^{-s}}.
\end{equation}
Using Lemma \ref{lemfdr}, we have $\Fnrb(\mnr)=\grando{n/k_n^{1+\alpha}}$ 
leading to
\begin{equation}
\Abs{W_n} =  \grando{\frac{n}{h_n^{1/2}k_n^{1/2+\alpha}}} + 
\grando{\frac{n^2}{h_n^{1/2}k_n^{3/2+2\alpha}}}+ \petito{\frac{k_n}{h_n n^s}}.
  \end{equation}
Therefore, choosing $s=2$ entails $W_n \to 0$  and  $\psi_{n,x}(t) \to e^{-t^2/2}$ as $n\to\infty$.
\CQFD

\section{Bias reduction}
\label{secred}

Theorem \ref{thnorasymp1} is not convenient for providing confidence intervals for $f(x)$,
since it only gives the asymptotic distribution of $\sigma_n^{-1}(\fnchapx-\E(\fnchapx))$.
A quick calculation shows that even in the case $f(x)\equiv a>0$, there is no hope of a similar
result for $\sigma_n^{-1}(\fnchapx-a)$. Indeed,
\begin{equation}
\E\left(a-\xnret\right)=\frac{k_n}{nc}\left(1-\exp{\left(-\frac{nca}{k_n}\right)}\right)
\geq \frac{1}{2}\frac{k_n}{nc}
\end{equation}
for $n$ large enough.
Thus, for every $x\in[0,1]$, under the hypotheses of Theorem \ref{thnorasymp1}:
\begin{equation}
\sigma_n^{-1}(a-\E(\fnchapx))=\sigma_n^{-1}\frac{1}{d_n}\sum_{x_r\in J_\ell(x)} 
(a-\E(\xnret)) 
 \geq  \frac{1}{2}\frac{nc d_n^{1/2}}{k_n}\frac{k_n}{nc}=\frac{1}{2}d_n^{1/2}\to \infty.
\end{equation}
Define $\check{f}_n(x)=\hat{f}_n(x)+k_n/(nc)$. Since $c$ is not supposed to be known,
$\check{f}_n$ is of no use in practice. Nevertheless, it is a first theoretical step
to introduce an estimator with a smaller bias and to study its asymptotic distribution
(see Theorem \ref{thnorasymp4}).
\begin{Theo}
\label{thnorasymp3}
If $f$ is $\alpha$-Lipschitzian $(0<\alpha\leq 1)$, under the three conditions 
$\Cz$, $\Cu$, and $n=\petito{k_n^{1/2} h_n^{1/2+\alpha}}$,
${\sigma}_n^{-1}(\check{f}_n(x) - f(x))$
converges in distribution to a standard Gaussian variable.
\end{Theo}
\proof
The proof is based on the expansion
\begin{equation}
\label{eqsuperdec}
\sigma_n^{-1}(\check{f}_n - f)= \sigma_n^{-1}(\check{f}_n - \E(\check{f}_n))+
\sigma_n^{-1}(\E(\check{f}_n)-f_n)+ \sigma_n^{-1}(f_n-f).
\end{equation}
Theorem \ref{thnorasymp1} shows that the first term converges in distribution to a standard Gaussian variable.
Now, from Lemma \ref{lemespe}~(\ref{deuxi}),
\begin{equation}
\max_{1\leq r\leq k_n}\Abs{\E\left(\xnret\right)+\frac{k_n}{nc}-k_n\lnr}=
\grando{\frac{1}{k_n^{\alpha}}},
\end{equation}
and we obtain,
\begin{equation}
\label{unibound}
\!\!\sigma_n^{-1}\left(\E(\check{f}_n(x)-f_n(x)\right)\!=
\sigma_n^{-1}(h_n+1)\!\!\! \sum_{x_r\in J_{\ell}(x)} 
\left(\frac{\E\left(\xnret\right)}{k_n}+\frac{1}{nc}-\lnr\right)=
\grando{\frac{n}{k_n^{1/2+\alpha}h_n^{1/2}}}.
\end{equation}
Finally, let us consider the third term. In view of Proposition~\ref{propcu},
\begin{equation}
\sigma_n^{-1}\Abs{f_n(x)-f(x)} =  \frac{ncd_n^{1/2}}{k_n}\grando{h_n^{-\alpha}}=
\grando{\frac{n}{k_n^{1/2} h_n^{1/2+\alpha}}}
\to 0.
\end{equation}
\CQFD
\noindent We are now ready to introduce a very simple ad hoc estimate of $k_n/(nc)$:
\begin{equation}
\label{defzn}
Z_n=\frac{1}{k_n}\somr \znret,
\end{equation}
where $\znret$ is defined by
$\znret=0$ if $N^{*n}(\Dnr)=0$ and
$\znret$ is the infimum of the second coordinates of the
points of the truncated process $N^{*n}(.\cap \Dnr)$ if $N^{*n}(\Dnr)>0$.
Define 
\begin{equation}
\tilde{f}_n(x)=\sum_{r=1}^{k_n} K_n(x_r,x)\left(\frac{\xnret+Z_n}{k_n}\right).
\end{equation}
\noindent The moments of $\znret$ can be expanded similarly to that of $\xnret$,
leading to
\begin{equation}
\label{moments}
\E(Z_n)= \frac{k_n}{nc} + \petito{n^{-s}},\;\forall s>0 \mbox{ and }
\Var(Z_n) \sim \frac{k_n}{n^2 c^2}.
\end{equation}
We now prove the following theorem which states that $\tilde{f}_n$ inherits 
its asymptotic behavior from $\check{f}_n$.
\begin{Theo}
\label{thnorasymp4}
Under conditions of Theorem \ref{thnorasymp3}, $\sigma_n^{-1}(\tilde{f}_n(x)-f(x))$
converges in distribution to a standard Gaussian random variable.
\end{Theo}
\noindent\proof
In view of Theorem \ref{thnorasymp3}, it suffices to verify that $\sigma_n^{-1}
(\check{f}_n(x)-\tilde{f}_n(x))$ converges to 0 in probability.
We shall show that for all $\varepsilon>0$,
\begin{equation}
\proba{\sigma_n^{-1} \Abs{\check{f}_n(x)-\tilde{f}_n(x)}>\varepsilon} \leq
\proba{\sigma_n^{-1} \Abs{ Z_n - \frac{k_n}{nc} } > \varepsilon} 
\end{equation}
converges to zero as $n$ goes to infinity. To this end, consider the expansion
\begin{equation}
\sigma_n^{-1} \Abs{ Z_n - \frac{k_n}{nc} } \leq \sigma_n^{-1} \Abs{ Z_n - \E(Z_n) }
+ \sigma_n^{-1} \Abs{ \E(Z_n) - \frac{k_n}{nc} }.
\end{equation}
In view of (\ref{moments}), the second term is bounded from above by $\varepsilon/2$ for
$n$ large enough, and therefore
\begin{eqnarray}
\proba{\sigma_n^{-1} \Abs{\check{f}_n(x)-\tilde{f}_n(x)}>\varepsilon} & \leq &
\proba{ \sigma_n^{-1} \Abs{ Z_n - \E(Z_n) } > \varepsilon/2 } \\
& \leq & \frac{4}{\varepsilon^2} \Var{  (\sigma_n^{-1} Z_n) } = \grando{1/h_n},
\end{eqnarray}
with (\ref{moments}).
\CQFD
\noindent
For instance, in the case where $f$ is continuously differentiable ($\alpha=1$),
one can choose $h_n=n^{1/2}$ and $k_n=n^{1/2}(\ln n)^{\varepsilon}$ for
all $\varepsilon>0$ arbitrarily small. Theorem \ref{thnorasymp4} shows
that $\tilde{f}_n(x)$  converges to $f(x)$ with the rate $n^{-1/2}(\ln n)^{\varepsilon}$ for all $x\in[0,1]$.

\section{Conclusion and further developments}

In this paper, the underlying Poisson point process is supposed to be homogeneous.
It is easily seen that all convergence theorems of section 4 still hold if the
intensity is an arbitary function on $S$, bounded from below by a constant $c>0$.
Such an extension for limit distributions is not straightforward.
We published a companion paper devoted to $C^1$-basis estimate.
This study, based on formula (\ref{difference}), can be
found in Girard and Jacob (2003b).
This case involves rather different difficulties due to the Dirichlet
kernel feature. However, this provides smooth estimators more adapted
to regular boundaries.

\section*{References}

\noindent  {Abbar, H.} (1990) 
{ Un estimateur spline du contour d'une r\'epartition ponctuelle al\'eatoire.}
{\em Statistique et analyse des donn\'ees}, {\bf 15}(3), 1--19.\\
\noindent { Abbar, H. and Suquet, Ch. } (1993)
{ Estimation $L^2$ du contour d'un processus de Poisson homog\`ene sur le plan.}
{\em Publ. IRMA Lille}, {\bf 31}, {\bf II}.\\
\noindent { Cencov, N. N.} (1962) 
{ Evaluation of an unknown distribution density from observations.}
{\em Soviet Math.} {\bf 3},  1559--1562.\\
\noindent{ Chevalier, J.} (1976)
{ Estimation du support et du contour d'une loi de probabilit\'e.}
{\em Ann. Inst. H. Poincar\'e Probab. Statist.}, sect. B, 12, 339--364.\\
\noindent  { de Haan, L. and Resnick, S.} (1994)
{ Estimating the home range.}
{\em J. Appl. Probab.}, {\bf 31}, 700--720.\\
\noindent { Deprins, D., Simar, L. and Tulkens, H.} (1984)
{ Measuring Labor Efficiency in Post Offices.} in 
{\em The Performance of Public Enterprises: Concepts and Measurements} by 
M. Marchand, P. Pestieau and H. Tulkens, North Holland ed, Amsterdam.\\
\noindent { Gayraud, G.} (1997)
{ Estimation of functionals of density support.}
{\em Math. Methods Statist.}  {\bf 6}(1), 26--46.\\
\noindent { Geffroy, J.} (1964) 
{ Sur un probl\`eme d'estimation g\'eom\'etrique.}
{\em Publ. Inst. Statist. Univ. Paris},
 {\bf XIII}, 191--200.\\
\noindent { Gensbittel, M. H.} (1979) 
{\em Contribution \`a l'\'etude statistique de r\'epartitions ponctuelles
al\'eatoires. }
Th\`ese, Universit\'e Pierre et Marie Curie, Paris.\\
\noindent Girard, S. and Jacob, P. (2003)   Extreme
values and Haar series estimates of point processes boundaries.
\textit{Scandinavian Journal of Statistics}, \textbf{30}(2), 369--384.\\
\noindent{Girard, S. and Jacob, P.} (2003b)
{Projection estimates of point processes boundaries.}
{\em Journal of Statistical Planning and Inference}, \textbf{116}(1), 1--15.\\
\noindent {H\"ardle, W., Park, B. U. and Tsybakov, A. B.} (1995a)
{ Estimation of a non sharp support boundaries.}
{\em J. Multivariate Anal.}, {\bf 43}, 205--218.\\
\noindent { H\"ardle, W., Hall, P. and Simar, L.} (1995b)
{ Iterated boostrap with application to frontier models.}
{\em J. Productivity Anal.}, {\bf 6}, 63--76.\\
\noindent { Jacob, P.} (1984) 
{ Estimation du contour discontinu d'un processus ponctuel sur le plan.}
{\em Publ. Inst. Statist. Univ. Paris},
{\bf XXIX}, 1--25.\\
\noindent { Jacob, P. and Abbar, H.} (1989) 
{ Estimating the edge of a Cox process area.}
{\em Cahiers du Centre d'Etudes de Recherche Op\'erationnelle}, {\bf 31}, 215--226.\\
\noindent { Jacob, P. and Suquet, P.} (1995)
{ Estimating the edge of a Poisson process by orthogonal series.}
{\em J. Statist. Plann. Inference}, {\bf 46},
 215--234.\\
\noindent{ Korostelev, A., Simar, L. and Tsybakov, A. B.} (1995)
{ Efficient estimation of monotone boundaries.}
{\em Ann. Statist.}, {\bf 23}, 476--489.\\
\noindent{  Korostelev, A. P. and Tsybakov, A. B.} (1993)
{ Minimax linewise algorithm for image reconstruction.}
in {\em Computer intensive methods in statistics} by W. H\"ardle, L. Simar ed.,
Statistics and Computing, Physica Verlag, Springer.\\
\noindent{ Mammen, E. and Tsybakov, A. B.} (1995)
{ Asymptotical minimax recovery of set with smooth boundaries.}
{\em Ann. Statist.}, {\bf 23}(2), 502--524.\\
\noindent{ Tiago de Oliveira, J.} (1963) 
{ Estatistica de densidades;~Resultados assintoticos.}
{\em Rev. Fac. Ci. Univ. Lisboa} {\bf A9}, 65--171.

\section*{Authors' address}

\noindent
St\'ephane Girard and Pierre Jacob,
Laboratoire de Probabilit\'es et Statistique, Universit\'e Montpellier 2,
place Eug\`ene Bataillon, 34095 Montpellier cedex 5, France.
{\tt jacob@stat.math.univ-montp2.fr}

\section*{Appendix: Auxilary results}

In the following lemma we give 
the distribution function of $\xnret$ ($1\leq r\leq k_n$), with an accurate approximation determined by the bounds $\mnr$ and $\Mnr$ of $f$ on $\inr$
(the notations are those introduced in section \ref{prelimin}).
The proof is straightforward
after noticing that, for every measurable set $B$, $P(N^{*n}(B)=0)=\exp{(-nc\lambda(B))}$.

\begin{Lem}
\label{lemfdr}
The distribution function of $\xnret$ ($1\leq r\leq k_n$) is given by
\begin{equation}
\begin{array}{l}
\Fnr(u)=\\
\\
\\
\end{array}
\left|
\begin{array}{ll}
0 & \mbox{ if } u < 0, \\
\exp{\left[ \frac{nc}{k_n} (u-\Mnr)-nc\enr(u)\right]} & \mbox{ if } 0 \leq u \leq \Mnr, \\
1 & \mbox{ if } \Mnr \leq u, 
\end{array}
\right.
\end{equation}
with $\enr(u)=\lnr-\frac{\Mnr}{k_n}$, if $0 \leq u \leq \mnr$,
 and $\lnr-\frac{\Mnr}{k_n}\leq\enr(u)\leq 0$, if $\mnr \leq u \leq \Mnr$.
\end{Lem}
An approximation of the first moments can be derived from the distribution function given in Lemma \ref{lemfdr}:

\begin{Lem}
\label{lemespe}
Suppose $f$ is $\alpha$-Lipschitzian $(0<\alpha\leq 1)$ and $\Cu$. Then,
\begin{equation}
\label{uni}
\max_{1\leq r\leq k_n} \E\left(\ynr^2\right)=\grando{\frac{1}{n^2}} + \grando{\frac{1}{k_n^{2(1+\alpha)}}},
\end{equation}
and
\begin{equation}
\label{deuxi}
\max_{1\leq r\leq k_n}\Abs{\E\left(\xnret\right)-\anr}=\grando{\frac{1}{k_n^\alpha}},
\end{equation}
where $\anr=\mnr-\frac{k_n}{nc}$.  If moreover $\Cd$,
\begin{equation}
\label{troisi}
\Var\left(\xnret\right)\sim\frac{k_n^2}{n^2 c^2},\;\;  1\leq r\leq k_n.
\end{equation}
\end{Lem}

\noindent
\proof
\vspace*{-5mm}
\begin{itemize}
\item Observe that
\begin{eqnarray}
\E\left[\left(\xnret-k_n\lnr\right)^2\right] = k_n^2 \lnr^2 P(\xnret=0)\!\!\! & + & \!\!\!\int_0^{k_n\lnr}(u-k_n\lnr)^2 \Fnr'(u)du \nonumber\\
\!\!\!& + & \!\!\!\int_{k_n\lnr}^{\Mnr} (u-k_n\lnr)^2 \Fnr(du).
\end{eqnarray}
The distribution function $\Fnr$ is given in Lemma \ref{lemfdr} using
$\enr$, a non-decreasing function on $[0, \Mnr]$. Then,  
the derivative $\enr'$ exists almost everywhere and is non-negative. 
We thus obtain
\begin{equation}
\label{majderF}
\Fnr'(u)\leq \frac{nc}{k_n} \exp{\left[\frac{nc}{k_n}\left(u-k_n\lnr\right)\right]} \mbox{ a. e.}
\end{equation}
with equality for $0\leq u\leq\mnr$. The following bound may be found
quite easily:
\begin{equation}
\E\left[\left(\xnret-k_n\lnr\right)^2\right] \leq k_n^2 \lnr^2 e^{-nc\lnr} + \frac{2k_n^2}{n^2c^2}+(\Mnr-\mnr)^2,
\end{equation}
for $r=1,\dots,k_n$, and (\ref{uni}) is proved.
\item From Lemma \ref{lemfdr},
$$
\E\left(\xnret\right) = \int_0^{\infty}\Fnrb(u)du 
 = \int_0^{\mnr} \left[1-\exp{\left(\frac{nc}{k_n}\left(u-k_n\lnr\right)\right)}\right]du
+\int_{\mnr}^{\Mnr}\Fnrb(u)du,
$$
where the last integral is positive and less than $(\Mnr-\mnr)$. Therefore, for $r=1,\dots,k_n$,
\begin{equation}
\mnr+\frac{k_n}{nc}e^{-nc\lnr}\left(1-e^{nc\frac{\mnr}{k_n}}\right)\leq\E(\xnret)\leq 
\Mnr+\frac{k_n}{nc}e^{-nc\lnr}\left(1-e^{nc\frac{\mnr}{k_n}}\right),
\end{equation}
and consequently
\begin{equation}
\Abs{\E\left(\xnret\right)-\anr}\leq (\Mnr-\mnr) + \frac{k_n}{nc} 
\Abs{1+e^{-nc\lnr} - e^{-nc(\lnr-\mnr/k_n)}}.
\end{equation}
The first term of the sum is obviously $\grando{1/k_n^{\alpha}}$, uniformly in $r$. Furthermore
\begin{eqnarray}
\!\!\!\!\!\!\!\!\!\!\!\!\Abs{1+e^{-nc\lnr} - e^{-nc(\lnr-\mnr/k_n)}}\!\!\!\! &\leq &\!\!\!\! e^{-nc\mnr/k_n} + 1 - e^{-nc(\Mnr-\mnr)/k_n}\nonumber\\
\!\!\!\!& \leq & \!\!\!\!e^{-nc\mnr/k_n} + \frac{nc(\Mnr-\mnr)}{k_n} 
 =  \grando{\frac{n}{k_n^{1+\alpha}}},
\end{eqnarray}
uniformly in $r$ and (\ref{deuxi}) is proved.
\item (\ref{troisi}) is established with similar techniques.
\CQFD
\end{itemize}
The expression of the characteristic function can also be deduced from the distribution function given in Lemma \ref{lemfdr}:

\begin{Lem} 
\label{lemphi}
Suppose $f$ is $\alpha$-Lipschitzian $(0<\alpha\leq 1)$ and $\Cu$.
Let $(t_n)\in{\mathbb R}$ be a sequence such that $t_n=\petito{n/k_n}$ and
$t_n=\petito{k_n^{1+2\alpha}/n}$.
Then, the characteristic function of $(\xnret-\mnr)$ can be written at point $t_n$ as:
$$
\phinr(t_n)= \frac{1 + it_n \frac{k_n}{nc} \Fnrb(\mnr) +
\grando{\Abs{t_n}\frac{n}{k_n^{1+2\alpha}}} + \petito{n^{-s}}}{1+ it_n \frac{k_n}{nc}},
$$
with $s$ arbitrary large, and $\Fnrb=1-\Fnr$ ($1\leq r\leq k_n$).
\end{Lem}
\proof
Consider the expansion
\begin{equation}
\E(e^{i t_n \xnret})=P(\xnret=0)+\int_0^{\mnr}e^{ix t_n}\Fnr'(x)dx + \int_{\mnr}^{\Mnr}
 e^{ixt_n} \Fnr(dx).
\end{equation}
The first and second term can be computed explicitely since Lemma \ref{lemfdr} provides
an expression of $\Fnr$ on $[0,\mnr]$:
\begin{equation}
\label{tmp1}
\E(e^{i t_n\xnret})=e^{-nc\lnr} + \frac{ e^{i t_n \mnr} \Fnr(\mnr) - e^{-nc\lnr}}
{1 + i t_n \frac{k_n}{nc}} + \int_{\mnr}^{\Mnr} e^{ixt_n} \Fnr(dx).
\end{equation}
The third term expands as
\begin{equation}
\label{tmp2}
\int_{\mnr}^{\Mnr} e^{ixt_n} \Fnr(dx)= e^{i t_n\mnr}\Fnrb(\mnr)
+\int_{\mnr}^{\Mnr} (e^{ixt_n}-e^{it_n\mnr})\Fnr(dx),
\end{equation}
with $\Abs{e^{ixt_n}-e^{it_n\mnr}}\leq (\Mnr-\mnr)\Abs{t_n}$. Thus
\begin{eqnarray}
\Abs{\int_{\mnr}^{\Mnr} (e^{ixt_n}-e^{it_n\mnr})\Fnr(dx)}
& \leq & (\Mnr-\mnr)\Abs{t_n}\Fnrb(\mnr) \nonumber \\
& \leq & (\Mnr-\mnr)\Abs{t_n} \frac{nc}{k_n}(k_n\lnr-\mnr)\nonumber \\
\label{tmp3}
& = & \grando{\Abs{t_n}\frac{n}{k_n^{1+2\alpha}}}.
\end{eqnarray}
Now, take the common denominator in (\ref{tmp1}) and replace (\ref{tmp2})
and (\ref{tmp3}) in the numerator.
Notice that $k_n=\petito{{n}/{\ln n}}$
yields $e^{-nc\lnr}=\petito{n^{-s}}$, $\forall s>0$ and
multiply everything by $e^{-it_n\mnr}$ to conclude the proof.
\CQFD
\noindent In order to be more self-contained, we quote here Lemma 4.1 of 
Jacob and Abbar~(1989) which is used to prove Theorem \ref{thweib}(ii).
Let us recall that a triangular array $\{\eta_{n,r},\;r=1,\dots,k_n\}$
is called a null array if $\displaystyle\max_{1\leq r\leq k_n} \Abs{\eta_{n,r}} \to 0$ as $n\to\infty$.
\begin{Lem}
\label{lemmnull}
Let $\{x_{n,r},\;r=1,\dots,k_n\}$ be a positive triangular array such that
$\{z_{n,r}\}=\{e^{-x_{n,r}}\}$ is a null array. 
If $\displaystyle \prod_{r=1}^{k_n}(1-z_{n,r})$ has a limit $\ell>0$, then
$\displaystyle \prod_{r=1}^{k_n}(1-z_{n,r}e^{\eta_{n,r}})$ has the same
limit $\ell$, whenever $\{\eta_{n,r}\}$ is a null array.
\end{Lem}

\end{document}